\documentclass[prl,aps,showpacs,groupedaddress,superscriptaddress,twocolumn]{revtex4-2}

\usepackage{amsmath,amssymb}
\usepackage[colorlinks=true,linkcolor=blue,urlcolor=blue,citecolor=blue]{hyperref}
\usepackage{nicefrac}
\usepackage{braket}
\usepackage{orcidlink}

\begin{document}
\title{Absence of gapless Majorana edge modes in few-leg bosonic flux ladders}

\author{F.~A.~Palm\orcidlink{0000-0001-5774-5546}}
\affiliation{Department of Physics and Arnold Sommerfeld Center for Theoretical Physics (ASC), Ludwig-Maximilians-Universit\"at M\"unchen, Theresienstr. 37, D-80333 M\"unchen, Germany}
\affiliation{Munich Center for Quantum Science and Technology (MCQST), Schellingstr. 4, D-80799 M\"unchen, Germany}
\affiliation{CENOLI, Universit\'e Libre de Bruxelles, CP 231, Campus Plaine, B-1050 Brussels, Belgium}
\author{C.~Repellin}
\affiliation{Univ. Grenoble-Alpes, CNRS, LPMMC, 38000 Grenoble, France}
\author{N.~Goldman}
\affiliation{CENOLI, Universit\'e Libre de Bruxelles, CP 231, Campus Plaine, B-1050 Brussels, Belgium}
\affiliation{Laboratoire Kastler Brossel, Coll\`ege de France, CNRS, ENS-PSL University, Sorbonne Universit\'e, 11 Place Marcelin Berthelot, 75005 Paris, France}
\author{F.~Grusdt\orcidlink{0000-0003-3531-8089}}
\affiliation{Department of Physics and Arnold Sommerfeld Center for Theoretical Physics (ASC), Ludwig-Maximilians-Universit\"at M\"unchen, Theresienstr. 37, D-80333 M\"unchen, Germany}
\affiliation{Munich Center for Quantum Science and Technology (MCQST), Schellingstr. 4, D-80799 M\"unchen, Germany}

\date{\today}

\begin{abstract}
    The search for Majorana excitations has seen tremendous efforts in recent years, ultimately aiming for their individual controllability in future topological quantum computers.
    A promising framework to realize such exotic Majorana fermions are topologically ordered non-Abelian phases of matter, such as certain fractional quantum Hall states.
	Quantum simulators provide unprecedented controllability and versatility to investigate such states, and developing experimentally feasible schemes to realize and identify them is of immediate relevance.
	Motivated by recent experiments, we consider bosons on coupled chains, subjected to a magnetic flux and experiencing Hubbard repulsion.
    At magnetic filling factor $\nu\!=\!1$, similar systems on cylinders have been found to host the non-Abelian Moore-Read Pfaffian state in the bulk.
    Here, we address the question whether more realistic few-leg ladders can host this exotic state and its chiral Majorana edge states.
	To this end, we perform extensive DMRG simulations and determine the central charge of the ground state.
    While we do not find any evidence of gapless Majorana edge modes in systems of up to six legs,  exact diagonalization of small systems reveals evidence for the Pfaffian state in the entanglement structure.
    By systematically varying the number of legs and monitoring the appearance and disappearance of this signal, our work highlights the importance of finite-size effects for the realization of exotic states in experimentally realistic systems.
\end{abstract}

\maketitle

\begin{figure*}
\centering
    \includegraphics{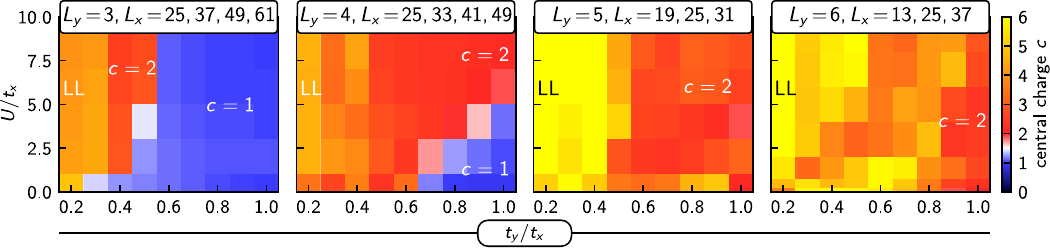}
    \caption{
        Central charge $c$ as extracted from the bipartite entanglement entropy using a scaling collapse for at most $N_{\rm max}=2$ particles per site and different numbers of legs $L_y=3, \hdots, 6$.
		For $L_y = 3, 4$ we find a rich phase diagram, including phases with $c\approx 1$ and $c\approx 2$.
        For all system sizes we find an increase of the central charge to $c\approx L_y$ at weak inter-leg coupling $t_y/t_x \to 0$ (Luttinger liquid regime).
        In the three-leg system ($L_y=3$), the $c=1$-phase persists even for strong repulsion, while this phase does not form in wider systems ($L_y=5,6$) at all, where only the Luttinger liquid regime and the $c=2$-phase could be identified.
        No extended phase with chiral Majorana edge states (half-integer central charge) is identified.
        }
    \label{fig:CentralCharge:Nu1:Nmax2}
\end{figure*}

\textit{Introduction.---}
Topologically ordered phases of matter in two dimensions -- for example fractional quantum Hall~(FQH) states~\cite{Stormer1983,Laughlin1983} -- are famously characterized by their ability to host anyonic excitations~\cite{Halperin1984,Arovas1984}.
Even more interestingly, there exist non-Abelian states (e.g. the Moore-Read Pfaffian state~\cite{Moore1991}) exhibiting non-Abelian anyonic excitations for which braiding operations may not commute.
Such systems could be used to realize topologically protected qubits and might provide a route towards topological quantum computation~\cite{Kitaev2003}.
Not surprisingly, there has been significant effort in identifying and manipulating such states, ultimately aiming for coherent control over individual non-Abelian anyons and their braiding processes.

Cold atom quantum simulators provide a suitable platform to address this challenge due to their high degree of control and the possibility to use interferometric techniques to probe topological properties~\cite{Atala2013,Duca2014,Grusdt2016,Braiding}.
A particularly promising approach in this regard are cold atoms in optical lattices subjected to an artificial magnetic field~\cite{Aidelsburger2013,Miyake2013} and strong repulsion~\cite{Tai2017}, a setup in which recently a $\nicefrac{1}{2}$-Laughlin state of two bosons has been realized~\cite{Leonard2023a}.
Similarly, a Laughlin state of two photons has been realized in a circuit QED setup~\cite{Wang2024}.
Consequently, the question arises whether similar lattice models can be employed to realize non-Abelian states of matter in experiments.

Previous numerical studies using exact diagonalization on small tori~\cite{Sterdyniak2012} as well as the density-matrix renormalization-group (DMRG) method on finite-width cylinders~\cite{Palm2021,Boesl2022} predicted the emergence of the paradigmatic Pfaffian state.
Hard-core three-body repulsion was found to stabilize the Pfaffian state -- consistent with the continuum Pfaffian's parent Hamiltonian~\cite{Moore1991}--, but also finite two-body repulsion turned out to be sufficient~\cite{Sterdyniak2012,Palm2021,Boesl2022}.
Existing numerical studies considered local observables like the particle number density and its (on-site) correlations or the pumped charge upon flux insertion.
Furthermore, entanglement properties like the topological entanglement entropy~\cite{Kitaev2006,Levin2006} and entanglement spectra~\cite{Li2008,Sterdyniak2011} have been used to identify and characterize topologically ordered bulk states.

Previous theoretical studies used periodic boundary conditions along some directions.
In contrast, open boundary conditions should be considered in view of realistically predicting the properties of experimental settings~\cite{Repellin2020}.
In such settings, the edge of the system may host chiral edge modes, which shed light on the topological order in the bulk.
For example, the edge spectrum of the Pfaffian state consists of one bosonic charge mode and one neutral Majorana mode, both of which are gapless and chiral~\cite{Teo2014}.
The low-energy edge theory of a given state can be captured in the central charge $c$, which counts the number of gapless modes and provides a useful quantity to characterize different topological orders.
While bosonic and fermionic modes each contribute central charge $c\!=\!1$, emergent Majorana edge modes give $c\!=\!\nicefrac{1}{2}$, thus resulting in central charge $c_{\rm Pf}\!=\!\nicefrac{3}{2}$ for the Pfaffian state~\cite{Fendley2007}.

Here, we numerically investigate the question whether few-leg flux ladders with tunable inter-leg tunneling and Hubbard interaction can host the Pfaffian state with its characteristic edge modes.
In particular, we use DMRG simulations to find the ground state on extended strips and extract from it the central charge via the scaling of the entanglement entropy.
We vary the number of legs in the ladder to highlight the role of finite-size effects along the shorter direction as well as to connect to realistic experimental setups.
Furthermore, we first use hard-core three-body repulsion to further stabilize the potential Pfaffian state, before relaxing this constraint in the light of experimental considerations.
We complement our analysis by exact diagonalization of small systems, with a special emphasis on the particle entanglement spectrum.

We did not find evidence of a gapless Majorana edge mode for up to six legs, independently of the interactions involved.
Nevertheless, we found non-trivial phases characterized by $c\!=\!1$ and $2$, respectively.
For systems of three and four legs we observe a transition between these two phases by increasing the strength of the Hubbard repulsion.
We attribute the absence of the gapless Majorana edge mode to significant finite-size effects in the ladder systems studied here, which future experiments will have to overcome.
Despite the absence of the gapless Majorana edge mode, our exact diagonalization shows evidence for the Pfaffian state in a narrow regime in sufficiently isotropic systems (5 legs and more).
Further theoretical effort is needed to find a suitable, experimentally realistic setting unambiguously hosting the Pfaffian state and its chiral Majorana edge states~\cite{Liu2024}.

\textit{Absence of gapless Majorana edge modes.---}
We study bosons on a ladder of $L_y$ legs with length $L_x\!\gg\!L_y$ with anisotropic hopping $t_y/t_x\!\leq\!1$ to compensate the geometric imbalance and to connect to earlier studies of coupled quantum wires~\cite{Kane2002,Teo2014}.
To obtain interacting topological many-body states, we introduce an on-site Hubbard repulsion of strength $U/t_x \!>\! 0$ and subject the system to a perpendicular magnetic field, resulting in the Hofstadter-Bose-Hubbard model, for which the Hamiltonian in Landau gauge reads
\begin{equation}
	\begin{aligned}
		\hat{\mathcal{H}} = &- t_x \sum_{x=1}^{L_x-1} \sum_{y=1}^{L_y} \left(\hat{a}^{\dagger}_{x+1, y}\hat{a}^{\vphantom{\dagger}}_{x,y} + \mathrm{H.c.}\right)\\
		&- t_y \sum_{x=1}^{L_x} \sum_{y=1}^{L_y-1} \left(\mathrm{e}^{i2\pi \alpha x}\hat{a}^{\dagger}_{x,y+1}\hat{a}^{\vphantom{\dagger}}_{x,y} + \mathrm{H.c.}\right)\\
		&+\frac{U}{2}\sum_{x,y} \hat{n}_{x,y}\left(\hat{n}_{x,y}-1\right).
	\end{aligned}
\label{Eq:HBH-Hamiltonian}
\end{equation}
Here, $\hat{a}^{(\dagger)}_{x,y}$ are bosonic annihilation (creation) operators and $\hat{n}_{x,y}\!=\!\hat{a}_{x,y}^{\dagger}\hat{a}^{\vphantom{\dagger}}_{x,y}$ are the boson number operators.
We study a strip geometry with open boundary conditions in both directions and work at magnetic flux $\alpha \!=\! N_{\phi} / \left[\left(L_x-1\right)\left(L_y-1\right)\right] \!=\! \nicefrac{1}{6}$ per plaquette -- close to the continuum limit~\cite{Soerensen2005,Hafezi2007} -- and magnetic filling factor $\nu\!=\!\nicefrac{N}{N_{\phi}}=1$.
At first, we also introduce a hard-core three-body repulsion by truncating the local Hilbert space to at most $N_{\rm max}\!=\!2$ bosons per site.

\begin{figure}[t]
    \centering
    \includegraphics{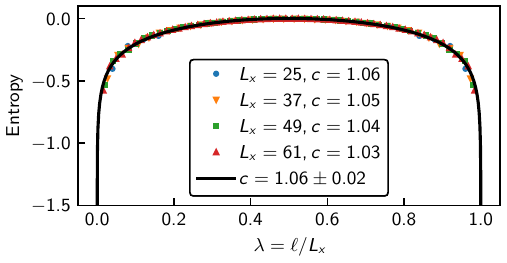}
    \caption{
        Bipartite entanglement entropy extracted from MPS simulations for a representative set of parameters ($U/t_x\!=\!2.0$, $t_y/t_x\!=\!1.0$) of a three-leg system, $L_y\!=\!3$.
        We find a clear scaling collapse of the numerical data.
		The fit with the CFT prediction $S_{\rm CFT}(\lambda)$ (black solid line) is performed using all data points in the bulk, leaving out the outermost 2 data points for each length $L_x$ on both ends of the system.
    }
    \label{fig:ScalingCollapse}
\end{figure}

We perform DMRG simulations~\cite{White1992,Schollwoeck2011,Hubig2015,HubigSyTen} to find an accurate matrix product state (MPS) representation of the ground state.
We exploit the $\mathrm{U}(1)$-symmetry associated with particle number conservation and consider bond dimensions up to $\chi=6000$~\cite{supp}.
From the MPS, we extract the bipartite entanglement entropy $S(\ell)$ as a function of the cut position $\ell$ along the long $x$-direction of the system.
The entanglement entropy $S(\ell)$ is related to the central charge $c$ via Calabrese~and~Cardy's CFT prediction~\cite{Calabrese2004},
\begin{equation}
	S_{\rm CFT}(\ell) = \frac{c}{6} \log\left(\frac{2L_x}{\pi} \sin\left(\frac{\pi \ell}{L_x}\right)\right) + g.
\end{equation}
Here, $g$ is some non-universal constant and $L_x$ is the length of the system.
This form can be rewritten as 
\begin{equation}
	S_{\rm CFT}(\lambda) = \frac{c}{6} \log\left(\sin\left(\pi\lambda\right)\right) + \mathrm{const.},
\end{equation}
where we introduced $\lambda \!=\! \nicefrac{\ell}{L_x}$ so that the non-constant part is independent of the system size and a scaling collapse is expected when the CFT prediction applies.
For a broad range of parameters, we find a good scaling collapse of the numerically obtained entanglement entropies~\cite{supp}.
We extract the central charge from a fit of the CFT prediction $S_{\rm CFT}(\lambda)$ to the complete collapsed set.
The extracted values of the central charge for a varying number of legs are given in Fig.~\ref{fig:CentralCharge:Nu1:Nmax2}.
An example for the scaling collapse for a representative set of parameters in a three-leg system is shown in Fig.~\ref{fig:ScalingCollapse}.

We observe a significant dependence on the number of legs $L_y$, but do not find an extended region with central charge $c\!=\!\nicefrac{3}{2}$ for $L_y\!=\!3,\hdots,6$ at any strength of the Hubbard interaction $U$.
While we cannot rule out a very narrow $c\!=\!\nicefrac{3}{2}$ region between the $c\!=\!1$ and $c\!=\!2$ phases, we have not found any signatures for a stable phase in this regime, featuring a quantized central charge of $c\!=\!\nicefrac{3}{2}$.
We conclude that there is no evidence for robust gapless Majorana edge modes in numerically accessible few-leg ladder systems of up to six legs.
This is in strong contrast to the case of the bosonic Laughlin state at $\nu\!=\!\nicefrac{1}{2}$, where the expected gapless chiral bosonic mode ($c\!=\!1$) already emerged in systems of $L_y \!\geq\! 3$ legs~\cite{Palm2022}.

Nevertheless, we find extended phases with central charges $c\!=\!1$ and $2$ close to the isotropic limit, $t_x\!\approx\! t_y$.
Three- and four-leg ladders host both the $c\!=\!1$ phase at weak interactions and the $c\!=\!2$ phase in the strong repulsion regime.
Note that in isotropic ($t_y \approx t_x$) three-leg systems the $c\!=\!1$ phase persists even at strong interactions.
Systems of five or six legs do only exhibit a ground state with $c\!=\!2$ for sufficiently large $t_y/t_x$.
For all system sizes, we find an increase of the central charge to $c\!\approx\! L_y$ as $t_y/t_x \!\to\! 0$~\cite{supp}.

One possible interpretation of our numerical findings originates from a field-theoretical ``coupled-wire'' approach, where analytical considerations predict different possible phases at filling factor $\nu\!=\!1$, in particular a strongly paired phase ($c_{\rm SP} \!=\! 1$), a bilayer phase ($c_{\rm B}\!=\!2$), and the Pfaffian phase ($c_{\rm Pf}\!=\!\nicefrac{3}{2}$)~\cite{Kane2002,Teo2014}.
In this picture, increasing the Hubbard repulsion might destabilize the paired state ($c_{\rm SP}\!=\!1$) in favor of the (unpaired) bilayer state ($c_{\rm B}\!=\!2$).
While the Pfaffian state is a viable candidate based on renormalization group arguments in the continuum~\cite{Teo2014}, we find its edge structure to be very sensitive to finite-size effects and microscopic details of the lattice model.
Similarly, the emergence of the other candidate states depends on the system size, where for $L_y\!=\!3, 4$ legs states with both $c\!=\!1$ and $2$ are realized.

Previous studies of bosonic flux ladders with up to three legs also identified phases with central charge $c\!=\!1$ and $2$, which showed characteristic features of Meissner and vortex (liquid) phases~\cite{Piraud2015,Kolley2015,Greschner2016,Petrescu2017,Chen2020}; some of these phases were also shown to display precursor signatures of a Laughlin state~\cite{Petrescu2017}.
Exploring the fate of these various quasi-1D phases as one further increases the number of legs, and determining their possible connection to bulk fractional quantum Hall states such as the Pfaffian or strongly-paired phases, appear as interesting perspectives in this context.

Finally, we interpret the increase of the central charge at weak inter-leg coupling, $t_y/t_x\to 0$, as a signature of the formation of weakly coupled Luttinger liquids in this limit with one bosonic mode in each leg, thus resulting in $c_{\rm LL}\!=\!L_y$.

\textit{Realistic Hubbard interactions.---}
So far, we have discussed the case of the Hofstadter-Hubbard model with additional hard-core three-body repulsion (at most $N_{\rm max}\!=\!2$ bosons per site), which is not easily realizable in cold atom experiments~\cite{Daley2009,Hafezi2014,Lee2018,Petiziol2021}.
However, the existence of the interesting states with $c\!=\!1$ and $2$ discussed above naturally leads to the question whether these states could be realized in near-term quantum simulators.
Therefore, we now drop the three-body repulsion and expand the local Hilbert space to include up to $N_{\rm max}\!=\!4$ bosons per site.
Given the diluteness of the systems studied, we believe this to be an accurate description also for systems without any Hilbert space constraint at all.
For simplicity, we restrict our further investigation to four-leg ladders, where we previously found two exotic phases in the isotropic limit.

\begin{figure}
	\centering
	\includegraphics{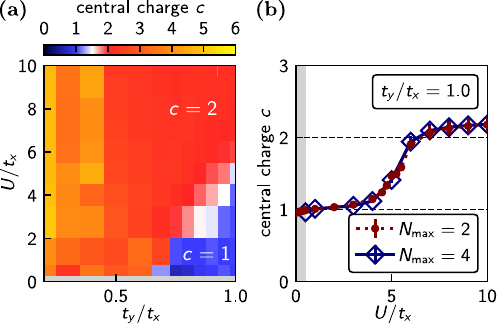}
	\caption{
		\textbf{(a)} Central charge $c$ as extracted from the bipartite entanglement entropy using a scaling collapse for $L_y\!=\!4$ and $N_{\rm max}\!=\!4$ without local three-body repulsion.
        We find the same phases as for the case $N_{\rm max}\!=\!2$ (see Fig.~\ref{fig:CentralCharge:Nu1:Nmax2}).
        \textbf{(b)} This is also evident in the isotropic limit, $t_y \!=\! t_x$, where the central charges for simulations with ($N_{\rm max} \!=\! 2$, red circles) and without ($N_{\rm max} \!=\! 4$, blue diamonds) hard-core three-body repulsion are in excellent agreement.
        In both panels the gray region indicates very weak interaction strengths $U/t_x < 0.5$ where for $N_{\rm max}\!=\!4$ the central charge could not be determined with the necessary accuracy.
        Ladder lengths used for the scaling collapse: $L_x\!=\!25, 33, 49, 61$.
	}
	\label{fig:CentralCharge:Nu1:Nmax4}
\end{figure}

Proceeding analogously to the previous study, we again extract the central charge for a wide range of parameters, see Fig.~\ref{fig:CentralCharge:Nu1:Nmax4}.
We find an essentially unchanged phase diagram with extended regions of $c\!=\!1$ and $2$ close to the isotropic limit.
For isotropic coupling $t_y\!=\!t_x$, we find excellent quantitative agreement of the central charge extracted for the Hilbert space truncations $N_{\rm max}=2$ and $4$.
We conclude that cold atom simulators should be able to realize and investigate these exotic states of matter already in systems of a few legs.

\textit{Evidence for the Pfaffian state in short ladders.---}
To complement our DMRG results, we have performed exact diagonalization calculations of the Hofstadter-Bose-Hubbard model in Eq.~\eqref{Eq:HBH-Hamiltonian} on short $L_x \times L_y$ ladders, with $L_y\!=\!2, \hdots, 7$ and $L_x > 7$.
To keep computing costs manageable, we have restricted our calculations to $N\!=\!4$ interacting bosons, and adjusted the value of $L_y$ and $L_x$ such that $48 \!\leq\! L_x L_y \!\leq\! 60$.

In such small systems the edge will exhibit strong finite-size effects, so that we do not necessarily expect to find gapless edge modes~\cite{Repellin2020}.
Instead, to analyze the nature of the many-body ground state, we have calculated its particle entanglement spectrum (PES) for a particle partition with $N_A\!=\!N_B\!=\!2$.
For a fractional quantum Hall ground state, the PES~\cite{Li2008, Sterdyniak2011} is gapped, and the number of states below the gap is equal to the degeneracy of quasihole states for a system with $N_A$ bosons in the same geometry, and is therefore a universal indicator of the nature of the ground state.
Being an integer by construction, the number of states below the gap is less prone to finite-size effects than other quantities obtained numerically, like the many-body Chern number~\cite{Niu1985} or the central charge.

\begin{figure}
	\centering
    \includegraphics{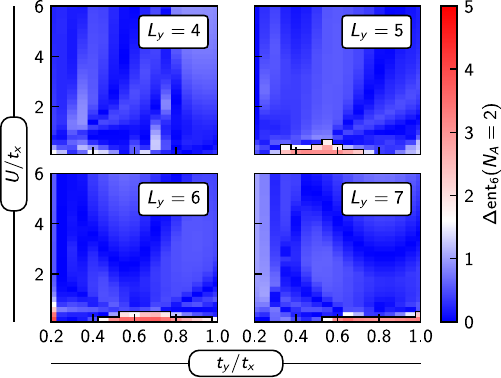}
	\caption{Stability regime of the bosonic Pfaffian state in short ladders, as extracted through the particle entanglement spectrum of 4-boson ground states obtained through exact diagonalization. The colormap indicates the entanglement gap above the lowest $6$ states for a particle cut with $N_A\!=\!2$. The red black lines indicates a finite entanglement gap with both particle cuts $N_A\!=\!1$ and $N_A\!=\!2$, with a counting consistent with the 4 particle Pfaffian (respectively 3 and 6 states below the entanglement gap). The ladder lengths (respectively $L_x\!=\!12, 10, 9, 8$) are adjusted to the number of legs $L_y$ to have a sufficient number of ladder sites while keeping the computational complexity manageable.
	}
	\label{fig:PES_ED}
\end{figure}

Our numerical results indicate that the bosonic Pfaffian state is absent in the narrow ladder geometry (\mbox{$L_y \!\leq\! 4$}), and emerges in more two-dimensional systems (\mbox{$L_y \!\geq\! 5$}) at small values of the Hubbard interaction $U/t_x \!\simeq\! 1$, see Fig.~\ref{fig:PES_ED}~\cite{supp}.
In the latter case, a moderate geometric anisotropy (relatively small number of legs $L_y$) may be compensated by adjusting the hopping anisotropy $t_y/t_x$~\cite{michenPRR2023}.
Our results also indicate that the specific value of both the ladder length $L_x$ and width $L_y$ affects the phase boundaries, but not the qualitative nature of each phase.

Finally, we have also performed the same calculations using a three-body on-site interaction instead of the two-body Hubbard interaction. Similarly, we find no evidence for a Pfaffian ground state in narrow ladders ($L_y\!\leq\!4$).
In general, the three-body repulsion favors a Pfaffian ground state at filling fraction $\nu\!=\!1$ of an isotropic Chern insulator model~\cite{supp,Sterdyniak2012}. 
This analysis thus confirms that the absence of a Pfaffian ground state in the narrow ladders originates from the system's geometry and edge effects rather than the interaction.

\textit{Conclusion and outlook.---}
We have found no evidence for a gapless Majorana edge mode and its half-integral central charge in few-leg flux ladders of up to six legs, even in a system mimicking the parent Hamiltonian of the $\nu\!=\!1$ bosonic Pfaffian state in the continuum.
Instead, we have found extended phases with central charge $c\!=\!1$ and $2$, the existence of which depends on the number of legs.
The microscopic details of these states -- which also persist under experimentally realistic conditions -- deserve a more detailed analysis and should be addressed in the future.
We speculate that these phases might be related to the strongly paired and bilayer phase predicted in a field-theoretical ``coupled-wire'' approach~\cite{Teo2014}.

We complemented our DMRG study with exact diagonalization of small systems, where the particle entanglement spectrum can give important insight into the ground state's nature.
Here, we confirmed the absence of the Pfaffian state in narrow ladders, while finding evidence for its emergence for $L_y\!\geq\! 5$ for weak Hubbard repulsion.
Our combined DMRG and exact diagonalization results highlight the finite-size dependency of topological signatures.
While bulk PES signatures appear in ladders with as few as $5$ legs, gapless Majorana edge modes, as detected through the central charge, do not emerge until at least $7$ legs.

Based on these findings, we believe that the gapless Majorana edge mode will eventually emerge in even wider ladders.
The fragility of the Majorana edge mode in narrow ladders should be contrasted with the substantially simpler bosonic Laughlin state, where already $L_y\!\geq\!3$ legs were sufficient to observe the expected bosonic edge mode.

The findings in this work are especially relevant in light of the current experimental abilities in cold atom quantum simulators~\cite{Leonard2023a}.
Using adiabatic state preparation protocols, few-leg ladders similar to those investigated here are within reach~\cite{He2017,michenPRR2023,Wang2024a,Palm2024,Blatz2024}.
Subsequently, quantum gas microscopy can be used to study the microscopic structure of competing topological states in a simple model.
Furthermore, studies of similar systems suggest that entanglement properties like the central charge might be visible from Fock basis snapshots~\cite{Petrescu2014,Lukin2019,Palm2022}.
There are also proposals to directly probe the entanglement Hamiltonian~\cite{Pichler2016,Dalmonte2018,Zache2022,Nardin2023,Joshi2023}, recently allowing for a measurement of the entanglement spectrum of a Chern insulator in cold atoms~\cite{Redon2023}.
Edge state spectroscopy protocols~\cite{GoldmanPRL2012, BinantiPRR2024} could potentially be adjusted to the ladder geometry; they would yield the conformal field theory counting, an indicator of the presence or absence of Majorana modes beyond the central charge.
The phases found here might also be distinguishable by using observables routinely available in quantum gas microscopes, like the density, bulk and edge currents, or the momentum distribution along the different legs.
Complementary, means to probe the emergent Majorana modes~\cite{Sun2023} need to be developed in order to unambiguously confirm their existence.

\begin{acknowledgments}
\textit{Acknowledgments.---}
The authors would like to thank F.~Binanti, T.~Blatz, A.~Bohrdt, M.~Greiner, M.~Greiter, J.~Kwan, J.~L\'eonard, H.~Schl\"omer, and R.~Wilke for fruitful discussions.
This research was funded by the Deutsche Forschungsgemeinschaft (DFG, German Research Foundation) via Research Unit FOR 2414 under project number 277974659, and under Germany's Excellence Strategy -- EXC-2111 -- 390814868. This project has received funding from the European Research Council (ERC) under the European Union’s Horizon 2020 research and innovation programm (Grant Agreement no 948141) -- ERC Starting Grant SimUcQuam. 
Work in Brussels is supported by the ERC Grant LATIS and the EOS project CHEQS. 
C.R. acknowledges support from ANR through
Grant No. ANR-22-CE30-0022-01.
\end{acknowledgments}

\bibliographystyle{apsrev4-2}
\bibliography{bibliography.bib}

\newpage
\widetext
\newpage
\begin{center}
	\textbf{\Large{Supplemental Material}}
\end{center}
\setcounter{equation}{0}
\setcounter{figure}{0}
\setcounter{table}{0}
\setcounter{section}{0}
\setcounter{page}{1}
\makeatletter
\renewcommand{\theequation}{S\arabic{equation}}
\renewcommand{\thefigure}{S\arabic{figure}}
\renewcommand{\thesection}{S\Roman{section}}
\renewcommand{\bibnumfmt}[1]{[S#1]}

\section{Technicalities regarding DMRG simulations}
\begin{figure}[h]
	\centering
	\includegraphics{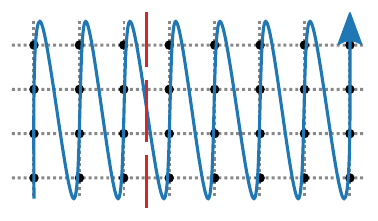}
	\caption{Given the lattice sites of the two-dimensional square lattice (black dots) the MPS chain (blue line) is chosen such that neighboring sites in $y$-direction are also neighboring sites in the MPS chain.
		Thus, a cut perpendicular to the legs in $x$-direction (red dashed line) can be realized by cutting a single bond in the MPS chain, giving immediate access to the bipartite entanglement entropy as function of the cut's position.}
	\label{Fig:Supp:MPS-Chain}
\end{figure}

Most of the numerical data presented in this work was obtained using the single-site variant of the density-matrix renormalization-group (DMRG) method~\cite{White1992,Schollwoeck2011} with subspace expansion~\cite{Hubig2015}, in particular using the SyTen implementation~\cite{HubigSyTen}.
We exploit the particle number $\mathrm{U}(1)$-symmetry to determine the variational ground state for a fixed number of particles.
Convergence of the simulations is ensured by comparing the ground state energy $\left\langle\hat{\cal{H}}\right\rangle$, the corresponding variance $\left\langle \hat{\cal{H}}^2 \right\rangle - \left\langle \hat{\cal{H}} \right\rangle^2$, and the entanglement entropy for spatial bipartitions for different bond dimensions up to $\chi\!=\!6000$.
The choice of the matrix product state (MPS) chain depicted in Fig.~\ref{Fig:Supp:MPS-Chain} allows for a simple evaluation of the bipartite entanglement entropy by cutting a single MPS bond.

\section{Extracting the central charge \& DMRG convergence}
As a function of the cut position $\ell$, the bipartite entanglement entropy $S(\ell)$ shows significant Friedel-like oscillations.
To account for these oscillations, we normalize the entanglement entropy in a given bond $\ell$ by the local densities at the sites $x\!=\!\ell\pm 1/2$~\cite{Palm2022},
\begin{equation}
	\tilde{S}(\ell) = \frac{2S(\ell)}{\left(n(\ell-\frac{1}{2}) + n(\ell+\frac{1}{2})\right)} \bar{n},
\end{equation}
with
\begin{equation}
	n(x) = \frac{1}{L_y} \sum_{y=1}^{L_y} \left\langle\hat{n}_{x,y}\right\rangle, \qquad \text{and} \qquad \bar{n} = \frac{1}{L_x}\sum_{x=1}^{L_x}n(x).
\end{equation}

In addition, to reduce the effect of the truncation at a finite bond dimension $\chi$, we extrapolate the rescaled entanglement entropies $\tilde{S}(\ell; \chi)$ to infinite bond dimensions as
\begin{equation}
	\tilde{S}(\ell; \chi) = \tilde{S}(\ell; \chi\to\infty) + A \log\left(1+\nicefrac{1}{\chi}\right).
\end{equation}

As discussed in the main text, the entanglement entropy is predicted to follow the scaling form~\cite{Calabrese2004}
\begin{equation}
	S_{\rm CFT}(\ell) = \frac{c}{6} \log\left(\frac{2L_x}{\pi} \sin\left(\frac{\pi\ell}{L_x}\right)\right) + g,
\end{equation}
which can be rewritten as
\begin{equation}
	S_{\rm CFT}(\lambda) = \frac{c}{6} \log\left(\sin\left(\pi\lambda\right)\right) + \text{const.} \text{ with } \lambda=\nicefrac{\ell}{L_x}.
	\label{eq:CFT_prediction_rescaled}
\end{equation}
We perform a scaling collapse of the rescaled entanglement entropies $\tilde{S}(\ell;\chi\to\infty)$ and extract the central charge $c$ from a fit using Eq.~\eqref{eq:CFT_prediction_rescaled}.

\begin{figure*}
	\centering
	\includegraphics{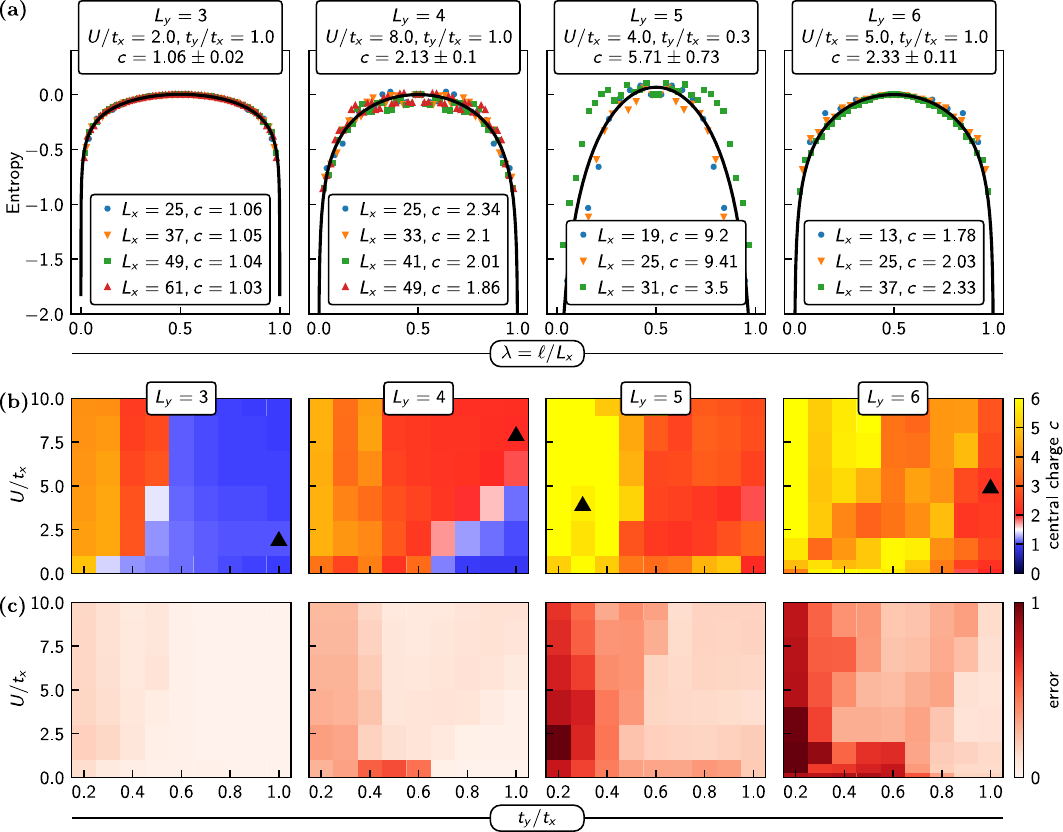}
	\caption{
		\textbf{(a)} Scaling collapse of the rescaled and shifted entanglement entropy extracted from MPS simulations for different parameters (indicated by the triangles in \textbf{(b)}) and varying numbers of chains $L_y$.
		For $t_y/t_x \!\approx\! 1$, we find a clear scaling collapse of the numerically extracted entanglement entropies, whereas in the numerically challenging weak coupling regime the collapse is less robust.
		The fit with the CFT prediction $S_{\rm CFT}(\lambda)$ (black solid line) is performed using all data points in the bulk, leaving out the outermost 2 data points for each system size on both ends of the chains.
		Central charge $c$ \textbf{(b)} and its fit error \textbf{(c)} as extracted from the bipartite entanglement entropy using a scaling collapse for at most $N_{\rm max}\!=\!2$ particles per site and different numbers of chains $L_y\!=\!3, \hdots, 6$.
		The various phases are discussed in the main text, see Fig.~1.
	}
	\label{fig:Supp:ScalingCollapse}
\end{figure*}

For the fit, we consider all data points in the bulk, omitting the two outermost points for each length $L_x$ on both ends of the ladders.
Furthermore, the fit error gives a lower bound on the error of the central charge.
We give examples for the scaling collapse, the extracted central charge, and the fit error in Fig.~\ref{fig:Supp:ScalingCollapse}.

\section{Central charge \& DMRG convergence in the weakly coupled limit}
While the scaling collapse and the subsequent fit work very well in the region close to the isotropic limit where $c\!=\!1$ and $2$, the weakly coupled regime $t_y/t_x \!\ll\! 1$ is more challenging.
In this regime, the DMRG calculations are difficult to converge so that the extracted central charge exhibits stronger numerical errors, see Fig.~\ref{fig:Supp:ScalingCollapse}.
Nevertheless, the significant increase beyond $c\!=\!2$ in the weakly coupled limit is consistent with the expected weakly coupled Luttinger liquids resulting in $c_{\rm LL}\!=\!L_y$.

\section{Particle entanglement spectra}
\begin{figure*}
	\centering
	\includegraphics{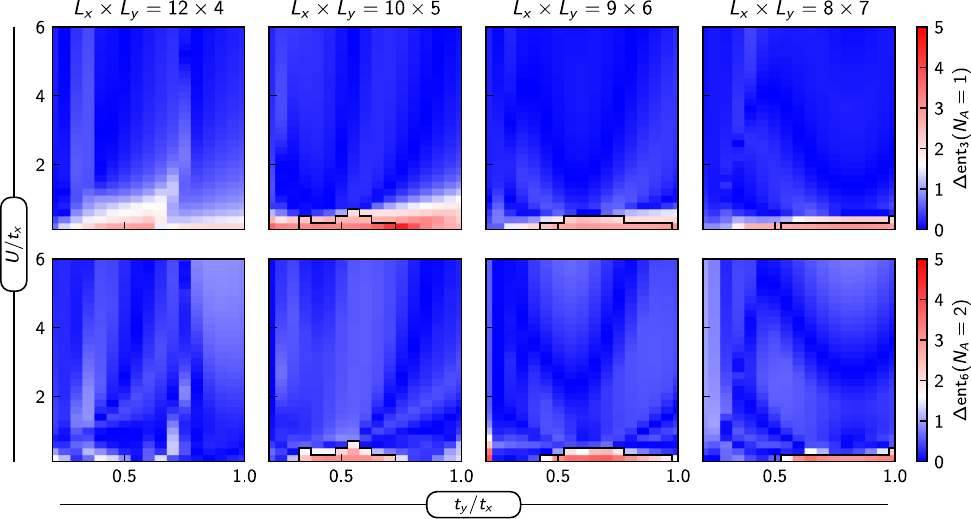}
	\caption{
		Entanglement gap above the lowest $3$ (upper row) and $6$ states (lower row) in the particle entanglement spectrum for $N_A\!=\!1$ and $2$ particles, respectively, resulting from tunable two-body interactions with strength $U$.
		The black line indicates the boundaries of the stability regions for the Pfaffian state.
	}
	\label{fig:Supp:EntanglementGap}
\end{figure*}

When identifying the stability region of the Pfaffian state in our exact diagonalization study, we rely on the particle entanglement spectrum (PES)~\cite{Li2008, Sterdyniak2011} as discussed in the main text.
In particular, we choose particle bipartitions with $N_A \!=\! 1$ and $2$ respectively.
For such cuts, the $N\!=\!4$ particle Pfaffian state is characterized by a sizable gap $\Delta\mathrm{ent}_{3}(N_A\!=\!1)$ ($\Delta\mathrm{ent}_{6}(N_A\!=\!2)$) in the PES above the lowest $3$ ($6$) states, respectively.
We show the entanglement gaps obtained from exact diagonalization in Fig.~\ref{fig:Supp:EntanglementGap}, which allows us to identify those parameters where the ground state exhibits the characteristics of the Pfaffian state as indicated in Fig.~4 in the main text.
We characterize the Pfaffian state by both entanglement gaps being larger than $1.6$.
Varying this cutoff slightly shifts the phase boundaries, but does not affect the qualitative features.

\subsection{Three-body repulsion}
\begin{figure*}
	\centering
	\includegraphics{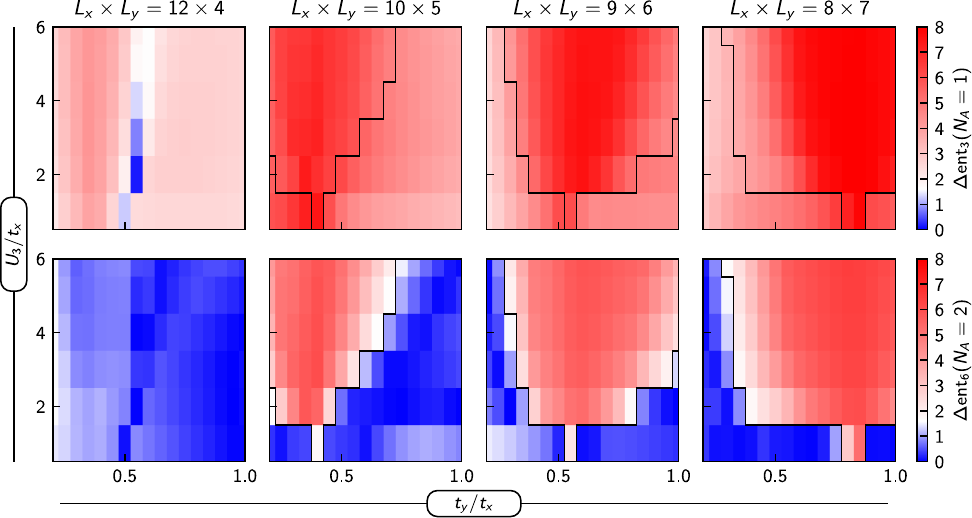}
	\caption{
		Entanglement gap above the lowest $3$ (upper row) and $6$ states (lower row) in the particle entanglement spectrum for $N_A\!=\!1$ and $2$ particles, respectively, resulting from tunable three-body interactions with strength $U_3/t_x$.
		The black line indicates the boundaries of the stability regions for the Pfaffian state.
	}
	\label{fig:Supp:EntanglementGap:ThreeBody}
\end{figure*}

We performed a similar analysis as before in the presence of three-body interactions of strength $U_3$, see Fig.~\ref{fig:Supp:EntanglementGap:ThreeBody}.
We find that in sufficiently isotropic systems, the finite three-body repulsion favors the Pfaffian state in agreement with earlier numerical studies~\cite{Sterdyniak2012}.

\bibliographystyle{apsrev4-2}
\bibliography{bibliography.bib}

\end{document}